\documentclass[fleqn,10pt]{wlscirep}
\title{Systemic Risk Analysis on Reconstructed Economic and Financial Networks}

\author[1,*]{Giulio Cimini}
\author[1]{Tiziano Squartini}
\author[2]{Diego Garlaschelli}
\author[1,3]{Andrea Gabrielli}
\affil[1]{Istituto dei Sistemi Complessi (ISC-CNR) UoS ``Sapienza'' Universit\`a di Roma, 00185 Rome, Italy}
\affil[2]{Lorentz Institute for Theoretical Physics, University of Leiden, 9506 Leiden, Netherlands}
\affil[3]{IMT Institute for Advanced Studies, 55100 Lucca, Italy}
\affil[*]{giulio.cimini@roma1.infn.it}

\begin{abstract}
We address a fundamental problem that is systematically encountered when modeling complex systems: the limitedness of the information available. 
In the case of economic and financial networks, privacy issues severely limit the information that can be accessed and, as a consequence, the possibility of correctly estimating 
the resilience of these systems to events such as financial shocks, crises and cascade failures. Here we present an innovative method to reconstruct the structure 
of such partially-accessible systems, based on the knowledge of intrinsic node-specific properties and of the number of connections of only a limited subset of nodes. 
This information is used to calibrate an inference procedure based on fundamental concepts derived from statistical physics, which allows to generate ensembles of directed weighted networks 
intended to represent the real system---so that the real network properties can be estimated with their average values within the ensemble. Here we test the method both on synthetic and empirical networks, 
focusing on the properties that are commonly used to measure systemic risk. Indeed, the method shows a remarkable robustness with respect to the limitedness of the information available, 
thus representing a valuable tool for gaining insights on privacy-protected economic and financial systems.
\end{abstract}
\begin{document}

\flushbottom
\maketitle

\thispagestyle{empty}

\section*{Introduction}

The estimation of the structural properties of a complex network when the available information on the system is incomplete represents an unsolved challenge \cite{Clauset2008,Mastromatteo2012}, 
yet it brings to many important applications. The most typical case is that of financial networks, whose nodes represent financial institutions and edges stand for financial ties 
(\emph{e.g.}, loans or derivative contracts)---the latter indicating dependencies among the institutions themselves, allowing for the propagation of financial distress across the network. 
The resilience of the system to the default or the distress of one or more institutions considerably depends on the topology of the whole network \cite{Battiston2012a,Battiston2012b,Foque2013};
however, because of confidentiality issues, the information on mutual exposures that regulators are able to collect is very limited \cite{Wells2004}.
Systemic risk analysis has been typically pursued by reconstructing the unknown links of the network using maximum entropy approaches \cite{Lelyveld2006,Degryse2007,Mistrulli2011}. 
These methods are also known as ``dense reconstruction'' techniques because they assume that the network is fully connected---an hypothesis that represents their strongest limitation. 
In fact, not only real networks show a largely heterogeneous distribution of the connectivity, but such a dense reconstruction was shown to lead to systemic risk underestimation 
\cite{Mastromatteo2012,Mistrulli2011}. More refined techniques like ``sparse reconstruction'' algorithms \cite{Mastromatteo2012} allow to obtain a network with arbitrary heterogeneity, 
however they still underestimate systemic risk because of the homogeneity principle used to assign link weights. 
A more recent approach \cite{Musmeci2013,Caldarelli2013} instead uses the limited topological information on the network to be reconstructed in order to generate an ensemble of graphs 
using the {\em configuration model} (CM) \cite{Park2004}---where, however, the Lagrange multipliers that define it are replaced by {\em fitnesses}, 
\emph{i.e.} known intrinsic node-specific features \cite{Caldarelli2002}. The average values of the observables computed on the CM-induced ensemble are then used as estimates for the real network properties. 
The latter approach overcomes the heterogeneity issue described above, yet it only allows to reconstruct systems in which each tie is undirected and unweighted---thus limiting the analysis 
to unrealistic and oversimplified configurations. Indeed, link directionality has been shown to play an important role in contagion processes and percolation analysis over these systems 
\cite{Boguna2005,Meyers2006} by, {\em e.g.}, speeding up or confining the infection with respect to the undirected case. Since real economic and financial networks are, by their nature, directed, 
links directionality has to be taken into account when assessing their robustness to shock and crashes. Moreover, the connection weights between the entities of these systems often assume heterogeneous values, 
which in turn strongly affect the way such entities react to the default or distress of their interacting partners \cite{Battiston2012b}. 

In order to achieve a realistic and faithful reconstruction of economic and financial networks, here we develop an improved procedure 
that allows to reconstruct links directionality, and at the same time we implement a simple yet effective prescription to assign link weights. 
Our method can thus be employed specifically for systemic risk estimation, by assessing those topological properties that have been shown to play a crucial role 
in contagion processes and in the propagation of distress over a network: the k-core structure \cite{Kitsak2010}, the percolation threshold \cite{Barrat2008}, the mean shortest path length \cite{Bellman1958} 
and the DebtRank \cite{Battiston2012b}. In particular, we perform an extensive analysis in order to quantify the accuracy of our method with respect to the size of the subset of nodes 
for which the topological information is available. Validation of the method is carried out on benchmark synthetic networks generated through a fitness-induced CM, 
as well as on two representative empirical systems, namely the International Trade Network (WTW) \cite{Gleditsch2002} and the (E-mid) Electronic Market for Interbank Deposits \cite{DeMasi2006}. 
In both cases, we have full information on these systems and we can thus unambiguously assess the accuracy of the method in describing them.

\section*{Method}

Before explaining our method in detail, let us introduce some notation. We will deal with weighted directed networks, {\em i.e.}, graphs composed by a set $V$ of nodes (with $|V|=N$) 
and described by a weighted directed adjacency matrix---whose generic element $w_{i\rightarrow j}$ represents the weight of the connection that runs from node $i$ to node $j$. 
The incoming total weight or {\em in-strength} for a generic node $i$ is then defined by $s_i^{in}=\sum_{j\in V} w_{j\rightarrow i}$, whereas, its outgoing total weight {\em out-strength} 
reads $s_i^{out}=\sum_{j\in V} w_{i\rightarrow j}$. It is also convenient to introduce the binary directed adjacency matrix that describes the binary topology: 
$a_{i\rightarrow j}=\Theta[w_{i\rightarrow j}]$ ($\Theta$ is the Heaviside step function: $\Theta[a]=1$ for $a>0$ and $\Theta[a]=0$ otherwise). 
This allows to define node $i$'s number of incoming connections or {\em in-degree} $k_i^{in}=\sum_{j\in V} a_{j\rightarrow i}$ 
and number of outgoing connections or {\em out-degree} $k_i^{out}=\sum_{j\in V}a_{i\rightarrow j}$. Finally, the binary undirected adjacency matrix---whose elements are obtained as 
$a_{ij}\equiv a_{ji}=\Theta[w_{i\rightarrow j}+w_{j\rightarrow i}]$---is used to define the number of incident connections or undirected {\em degree} of node $i$: $k_i=\sum_{j\in V}a_{ij}\equiv\sum_{j\in V}a_{ji}$. 

Given these ingredients, our network reconstruction method works as follows. Let us suppose to have incomplete information about the topology of a given network $G_0$. 
In particular, suppose to know the in-degree and out-degree sequences $\{k_i^{in}\}_{i\in I}$ and $\{k_i^{out}\}_{i\in I}$ only for a subset $I\subset V$ of all nodes (where $|I|=n<N$). 
Moreover, suppose to know a pair of intrinsic properties $\{\chi_i\}_{i\in V}$ and $\{\psi_i\}_{i\in V}$ for all the nodes---that will be our {\em fitnesses} (see below). 
The method then invokes a statistical procedure to find the most probable estimate for the value $X(G_0)$ of a given property $X$ computed on the network $G_0$, 
compatible with the aforementioned constraints. We build on two important hypotheses. 

I) The network $G_0$ is drawn from an ensemble $\Omega$ induced by a directed CM \cite{Squartini2011}---meaning that $\Omega$ is a set of networks that are maximally random, except for the ensemble averages 
 of the in/out-degrees $\{\langle k_i^{in}\rangle_\Omega\}_{i\in V}$ and $\{\langle k_i^{out}\rangle_\Omega\}_{i\in V}$ that are constrained to the observed values 
 $\{ k_i^{in} \}_{i\in V}$ and $\{ k_i^{out} \}_{i\in V}$, respectively \cite{Park2004}. 
 The directed CM prescribes that the probability distribution over $\Omega$ is defined via a set of Lagrange multipliers $\{x_i,y_i\}_{i\in V}$ (two for each node), 
 whose values can be adjusted in order to satisfy the equivalence $\langle k_i^{in} \rangle_\Omega \equiv k_i^{in}$ and $\langle k_i^{out} \rangle_\Omega \equiv k_i^{out}$, $\forall i\in V$ \cite{Squartini2011}. 
 The values of $x_i$ and $y_i$ are thus induced by the in- and out- degree of node $i$, respectively. The role of $\{x_i,y_i\}_{i\in V}$ in controlling the topology is better clarified 
 by writing explicitly the ensemble probability for a directed connection between any two nodes $i$ and $j$ \cite{Park2004}: 
 \begin{equation}
 p_{i\rightarrow j}\equiv\langle a_{i\rightarrow j}\rangle_\Omega=\frac{x_jy_i}{1+x_jy_i},\label{eq:prob linking dir}
 \end{equation}
 so that $x_i$ ($y_i$) quantifies the ability of node $i$ to receive incoming (form outgoing) connections. 
 
II) The fitnesses $\{\chi_i\}_{i\in V}$ and $\{\psi_i\}_{i\in V}$ are assumed to be linearly correlated, respectively, to the in-degree-induced and out-degree-induced Lagrange multipliers 
 $\{x_i\}_{i\in V}$ and $\{y_i\}_{i\in V}$ through universal (unknown) parameters $\alpha$ and $\beta$: $x_i\equiv\sqrt{\alpha}\chi_i$ and $y_i\equiv\sqrt{\beta}\psi_i$, $\forall i\in V$. 
 Therefore eq.~(\ref{eq:prob linking dir}) becomes: 
 \begin{equation}
 p_{i\rightarrow j}=\frac{\sqrt{\alpha}\chi_j\sqrt{\beta}\psi_i}{1+\sqrt{\alpha}\chi_j\sqrt{\beta}\psi_i}=\frac{z\chi_j\psi_i}{1+z\chi_j\psi_i},\label{eq:prob linking z dir}
 \end{equation}
 where we have defined $z\equiv\sqrt{\alpha\beta}$.
 Such an hypothesis is inspired by the so-called \textit{fitness} or \textit{hidden-variables} model \cite{Caldarelli2002}, which assumes the network topology to be determined by intrinsic properties 
 associated to each node of the network. Note that this approach has been already used in the past to model several empirical economic and financial networks \cite{DeMasi2006,Garlaschelli2004}, 
 possibly within the CM framework assuming a connection between fitnesses and Lagrange multipliers \cite{Garlaschelli2005}. 

These two hypotheses allow us to map the problem of evaluating $X(G_0)$ into the one of choosing the optimal CM ensemble $\Omega$ induced by the fitnesses $\{\chi_i\}_{i\in V}$ and $\{\psi_i\}_{i\in V}$, 
that is compatible with the constraints on $G_0$---given by the knowledge of $\{k_i^{in}\}_{i\in I}$ and $\{k_i^{out}\}_{i\in I}$. 
Indeed, because of the limited available information, finding the CM of the real system \cite{Park2004} is impossible, 
and we thus have to impose it by assigning {\em ad hoc} values to the Lagrange multipliers---whence the name of ``fitness-induced'' CM (FiCM). 
Once the FiCM ensemble $\Omega$ is determined (it is univocally defined by the set $\{x_i,y_i\}_{i\in V}$, and thus by the set $\{\chi_i,\psi_i\}_{i\in V}$), 
statistical mechanics of networks prescribes that the quantity $X(G_0)$ typically varies in the range $\langle X \rangle_{\Omega} \pm \sigma_X^{\Omega}$, 
where $\langle X \rangle_{\Omega}$ and $\sigma_X^{\Omega}$ are respectively average and standard deviation of property $X$ estimated over $\Omega$ \cite{Park2004}. We can thus use $\langle X \rangle_{\Omega}$ 
as a good estimation for $X(G_0)$. In practice, since we know the fitness values $\{\chi_i,\psi_i\}_{i\in V}$, in order to determine unambiguously the ensemble $\Omega$ 
we need to find the most likely value of the proportionality constant $z$ that defines $\Omega$ according to eq.~(\ref{eq:prob linking z dir}). 
This can be done using the partial knowledge of the degree sequences to estimate the appropriate value of $z$ through a maximum-likelihood argument \cite{Garlaschelli2004}, \emph{i.e.}, 
by comparing, for the nodes in the set $I$, the average number of incoming and outgoing connections in the ensemble $\Omega$ with their in-degrees and out-degrees observed in $G_0$: 
\begin{equation}\label{eq:estimatez dir}
\sum_{i \in I} \left[\langle k_i^{in} \rangle_\Omega + \langle k_i^{out} \rangle_\Omega\right] = \sum_{i \in I} \left[k_i^{in}+k_i^{out}\right].
\end{equation}
In the above expression, $\langle k_i^{in}\rangle_{\Omega}=\sum_{j(\neq i)}p_{j\rightarrow i}$ and $\langle k_i^{out}\rangle_{\Omega}=\sum_{j(\neq i)}p_{i\rightarrow j}$ 
contain the unknown parameter $z$ through eq.~(\ref{eq:prob linking z dir}), and since $\{\chi_i,\psi_i\}_{i\in V}$ and $\{k_i^{in},k_i^{out}\}_{i\in I}$ are known, eq.~(\ref{eq:estimatez dir}) defines 
an algebraic equation in $z$, whose solution allows to build the FiCM ensemble and, at the end, to obtain an estimation of $X(G_0)$---even with the knowledge of the in- and out- degree of just a single node. 

Summing up, the algorithm works as follows. Given a network $G_0$, two fitness values $\chi$ and $\psi$ for each of the $N$ nodes, and the in-degrees and out-degrees only for a subset $I$ of $|I|=n<N$ nodes:
\begin{itemize}
\item we compute the sum of the in-degrees and out-degrees of the nodes in $I$ %($\sum_{i \in I} \left[k_i^{in}+k_i^{out}\right]$) 
and use it together with the fitnesses $\{\chi_i,\psi_i\}_{i\in V}$ to obtain the value of $z$ by solving eq.~(\ref{eq:estimatez dir});
\item using such estimated $z$ and the fitnesses $\{\chi_i,\psi_i\}_{i\in V}$, we generate the ensemble $\Omega$ by placing a directed link from a given node $i$ to a given node $j$ with probability $p_{i\rightarrow j}$ 
of eq.~(\ref{eq:prob linking z dir});
\item we compute the estimate of $X(G_0)$ as $\langle X \rangle_{\Omega}\pm\sigma_X^{\Omega}$ in the FiCM ensemble, either analytically or numerically ({\em i.e.}, by measuring it on networks drawn from $\Omega$).
\end{itemize}

Note that the numerical generation of a sample network from $\Omega$ consists in building a binary directed adjacency matrix, so that its generic element $a_{i\rightarrow j}=1$ with probability 
$p_{i\rightarrow j}$ ({\em i.e.}, the existence probability for the link from $i$ to $j$ given by eq.~(\ref{eq:prob linking z dir})) and $a_{i\rightarrow j}=0$ otherwise. 
Finally, once the network is generated, in order to obtain a weighted topology we place $\forall i,j$ a weight $\tilde{w}_{i\rightarrow j}$ on the directed link from $i$ to $j$ 
(provided its existence, \emph{i.e.}, $a_{i\rightarrow j}=1$) according to the following prescription:
\begin{equation}
 \tilde{w}_{i\rightarrow j}=\frac{\chi_j\,\psi_i}{W\,p_{i\rightarrow j}}\,a_{i\rightarrow j}\equiv\frac{1}{W}(z^{-1}+\chi_j\,\psi_i)\,a_{i\rightarrow j}, \label{eq:weight z dir}
\end{equation}
where the last equality comes from eq.~(\ref{eq:prob linking z dir}). In this expression, the normalization $W$ represents the \emph{induced total weight of the network}, 
defined as the geometric mean of the sum of the fitnesses: $W=\sqrt{(\sum_i\chi_i)(\sum_i\psi_i)}$. 
Indeed, $W$ corresponds to the ensemble average of the total network weight: $W\equiv\sum_{ij}\langle\tilde{w}_{i\rightarrow j}\rangle_\Omega$, 
where $\langle\tilde{w}_{i\rightarrow j}\rangle_\Omega=(\chi_j\,\psi_i)\,\langle a_{i\rightarrow j}\rangle_\Omega/(W\,p_{i\rightarrow j})=(\chi_j\,\psi_i)/W$ 
is the ensemble average for the weight of the link from $i$ to $j$. 

\begin{figure}[ht]
\centering
\includegraphics[width=0.7\textwidth]{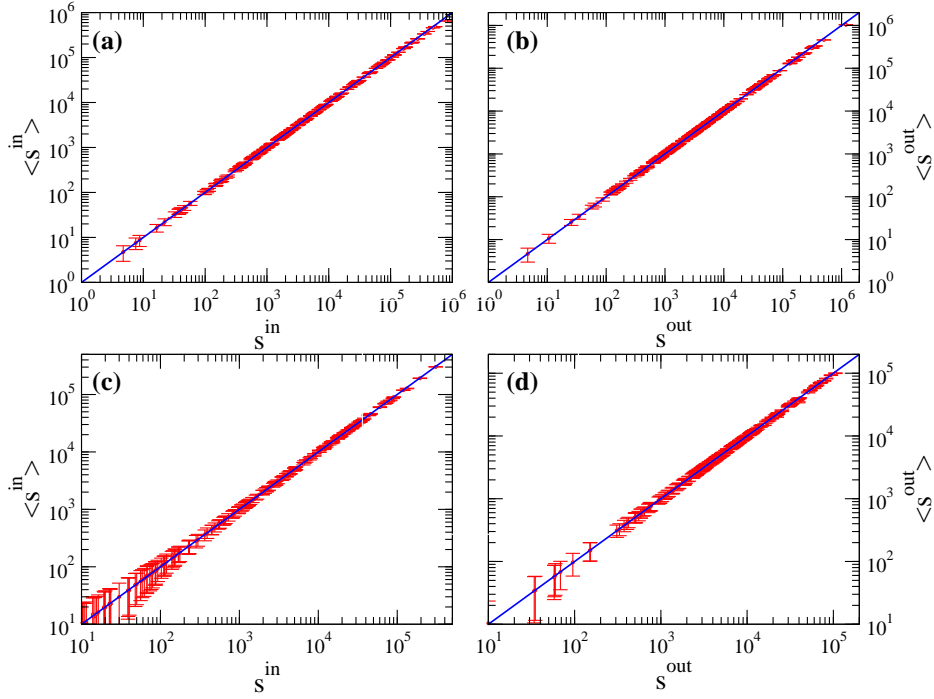}
\caption{Conservation of the strength sequences. Scatter plots of node in-strengths $s^{in}$ and out-strengths $s^{out}$ observed for the real network $G_0$ and their ensemble averages 
obtained from eq.~(\ref{eq:weight z dir}). Upper panels (a,b) refer to WTW, lower panels (c,d) to E-mid.}
\label{fig:test_check}
\end{figure}

Remarkably, thanks to this procedure to assign link weights, the ensemble averages of a node $i$'s total in-strength $\langle s_i^{in}\rangle_\Omega=\sum_{j\in V}\langle\tilde{w}_{j\rightarrow i}\rangle_\Omega$ 
and out-strength $\langle s_i^{out}\rangle_\Omega=\sum_{j\in V}\langle\tilde{w}_{i\rightarrow j}\rangle_\Omega$ turn out to be directly proportional to $\chi_i$ and $\psi_i$, respectively and $\forall i\in V$. 
This suggests a natural way for selecting appropriate quantities to play the role of fitnesses in our method: 
the straightforward interpretation for $\chi$ and $\psi$ is that of nodes in-strengths and out-strengths observed in the real network $G_0$. 
Indeed, the assumption $\chi_i=s_i^{in}$ and $\psi_i=s_i^{out}$, $\forall i\in V$, brings to $W=\sum_{i\in V}\chi_i\equiv\sum_{i\in V}\psi_i$, and finally to the important equivalences 
$\langle s_i^{in}\rangle_\Omega\equiv s_i^{in}$ and $\langle s_i^{out}\rangle_\Omega\equiv s_i^{out}$. This means that we successfully preserve, on average, the strength sequences of the real network $G_0$ 
(and thus its total weight). In other words, our network reconstruction method calibrated on in/out strengths is based on a null model 
constraining the in-degree and out-degree sequence of a subset of nodes, together with the relative in-strength and out-strength sequence (see Figure \ref{fig:test_check}). 

\section*{Empirical Dataset}

In order to test our network reconstruction method, we use two representative empirical systems of economic and financial nature. 
The first one is the international trade network of the World Trade Web (WTW) \cite{Gleditsch2002}, \emph{i.e.}, the network whose nodes are the countries 
and links represent trade volumes between them: thus, $w_{i\rightarrow j}$ is the monetary flux from country $i$ to country $j$ (the ``amount'' of the export from $j$ to $i$). 
The second one is the (E-mid) Electronic Market for Interbank Deposits \cite{DeMasi2006}: in this case, the nodes are banks 
and a link $w_{i\rightarrow j}$ from bank $i$ to bank $j$ represents the amount of the loan that $i$ granted to $j$. 

In the following analysis we will use and show results for WTW trade volume data of year 2000, and E-mid aggregated transaction data of year 1999 
(both temporal snapshots correspond to the largest size of the network). Analyses for other annual snapshots are reported in the Supplementary Information, and bring to comparable results.
In the light of the discussion at the end of the previous section, we will use as fitnesses $\chi_i$ ($\psi_i$) 
the real node in-strength $s_i^{in}=\sum_{j\in V} w_{j\rightarrow i}$ (out-strength $s_i^{out}=\sum_{j\in V} w_{i\rightarrow j}$), 
\emph{i.e.}, the total import (export) volumes of countries for WTW, and with the total liquidity borrowed (lent) by banks for E-mid.
Note that the goodness of any choice for the fitness values must be first validated according to hypothesis II of our method (see the first part of section Results).

\section*{Topological Properties}

As stated in the introduction, we will test our network reconstruction method focusing on the network properties (each playing the role of $X$ in the discussion of section Methods) 
which are commonly regarded as the most significant for describing the network resilience to systemic shocks and crashes. 
We first consider two properties defined for undirected networks (in order to reconstruct these properties, we use the undirected version of the method \cite{Musmeci2013}):
\begin{itemize}
\item Degree of the main core $k^{main}$ and size of the main core $S^{main}$, where a $k$-core is defined as the ``largest connected subgraph whose nodes all have at least $k$ connections" 
(within this subgraph), and the main core is the $k$-core with the highest possible degree ($k^{main}$) \cite{Dorogovtsev2010}. 
The main core is relevant to our analysis as it consists of the most influential spreaders (of, \emph{e.g.}, an infection or a shock) in a network \cite{Kitsak2010}.
\item Size of the giant component $S_{GC}$ at the bond percolation threshold $p^*=\bar{k}^{-1}$ ($\bar{k}$ is the mean degree of the network), 
where bond percolation is the process of occupying each link of the network with probability $p$, and $p^*$ is the critical value of $p$ at which a percolation cluster 
containing a finite fraction of all nodes first occurs \cite{Barrat2008}. 
Note that the percolation threshold at $p^*=\bar{k}^{-1}$ (that we take as reference value) is a feature proper of homogeneous graphs in the infinite volume limit, whereas, 
for scale-free networks in the same limit it is $p^*\rightarrow0$. Note also that a bond percolation process can be mapped into a SIR model with infection rate $\beta$ 
and uniform infection time $\tau$. In fact, by defining the trasmissibility $T=1-e^{-\tau\beta}$ as the probability that the infection will be transmitted from an infected node to 
at least a susceptible neighbor before recovery takes place, the set of nodes reached by a SIR epidemic outbreak originated from a single node is statistically equivalent 
to the cluster of the bond percolation problem (with $p\equiv T$) the initial node belongs to~\cite{Newman2002}. 
\end{itemize}

We then move to properties defined for directed graphs:
\begin{itemize}
\item Link reciprocity $r$, measuring the tendency of node pairs to form mutual connections. It is defined as the ratio between the number of bidirected links 
and the total number of network connections: 
$r=(\sum_{ij}a_{i\rightarrow j}a_{j\rightarrow i})/(\sum_{ij}a_{i\rightarrow j})$. Reciprocity is considered a sensible parameter for systemic risk, giving a measure of direct mutual exposure among nodes. 
\item Average shortest path length $\lambda$ \cite{Bellman1958}, where the shortest path length $\lambda_{i\rightarrow j}$ from node $i$ to node $j$ is the minimum number of links 
required to connect $i$ to $j$ (following link directions), and $\lambda=N(N-1)/(\sum_{i\neq j}\lambda_{i\rightarrow j}^{-1})$ (the harmonic mean is commonly used to avoid problems caused 
by pairs of nodes that are not reachable from one to another, and for which $\lambda$ diverges).
This quantity measures the number of steps that are required, on average, for a signal or a shock to propagate between any two nodes of the network.
\item The Group DebtRank $DR$ \cite{Battiston2012b}, a measure of the total economic value in the network that is potentially affected by a distress on all nodes amounting to $0<\Phi<1$, 
with $\Phi=1$ meaning default. In a nutshell, $DR$ is based on computing the recursive impact ({\em i.e.,} the reverberation on the network) of the initial distress, and is defined as: 
\begin{equation}\label{eq.DR}
DR=\sum_i (h_i^*-h_i^0)\nu_i
\end{equation}
where $h_i^*$ is the final amount of distress on $i$ ($h_i^0=\Phi$) and $\nu_i$ is the relative economic value of $i$. We refer to the original paper \cite{Battiston2012b} for the details on how to compute $DR$. 
\end{itemize}

\section*{Results}

\subsection*{Test of FiCM modeling}

When testing our network reconstruction procedure it is important to keep in mind that the method is subject to three different kind of errors. 
The first one comes from hypothesis I that the real network $G_0$ can be properly described by a CM, whose Lagrange multipliers are obtained by constraining the whole 
in-degree and out-degree sequences \cite{Park2004}. The second one derives instead from hypothesis II that the node fitnesses $\{\chi,\psi\}_{i\in V}$ are proportional to the CM's Lagrange multipliers $\{x,y\}_{i\in V}$, 
{\em i.e.}, from imposing a FiCM. Finally, the third one is due to the limited information available for calibrating the FiCM and obtain the true value of $z$---namely, 
the partial knowledge of the in-degree and out-degree sequences. 
Note however that the first source of mistakes cannot be controlled for in our context, as finding the CM that describes the data requires the knowledge 
of the whole in-degree and out-degree sequences (which is not accessible for our case studies). This is exactly why we have to make hypothesis II and impose a FiCM by assigning {\em ad hoc} values 
to the Lagrange multipliers. In this section we thus concentrate on the second source of errors. 

\begin{figure}[p]
\centering
\includegraphics[width=0.7\textwidth]{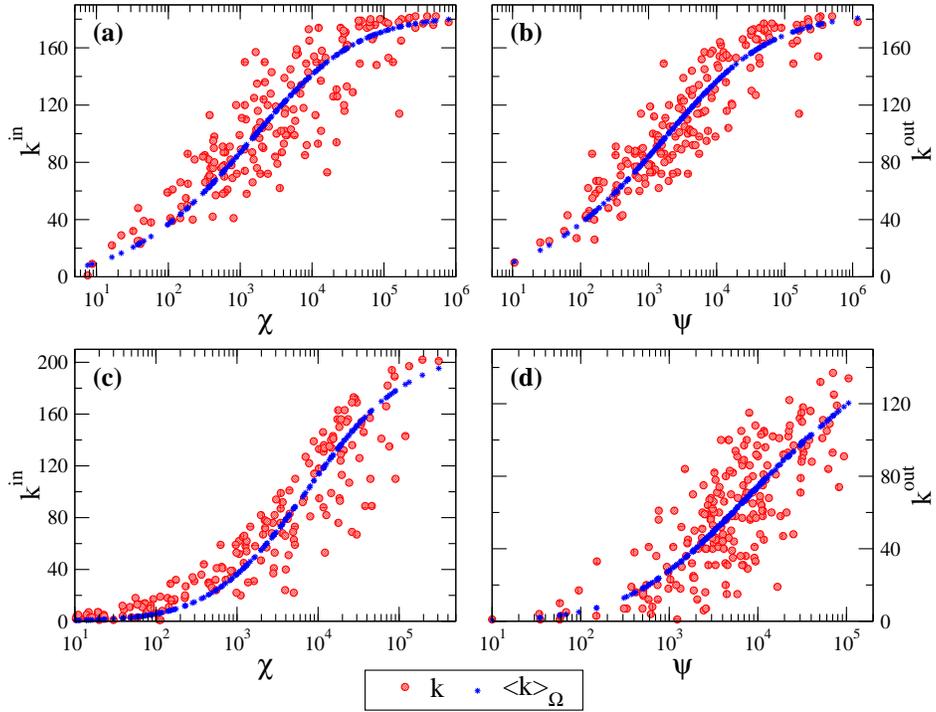}
\caption{Qualitative assessment for FiCM description of the real network $G_0$. 
Scatter plots of node fitnesses $\{\chi,\psi\}$ versus real node in- and out-degrees $\{k^{in},k^{out}\}$ of $G_0$ (red circles) and their ensemble averages computed via the FiCM 
(blue asterisks). Upper panels (a,b) refer to WTW, lower panels (c,d) to E-mid.}
\label{fig:test_fit}
\end{figure}

Indeed, real networks are not perfect realizations of the FiCM and can only be approximated by it \cite{Garlaschelli2004}. 
In order to assess qualitatively how well this FiCM describes the real network $G_0$, one can compare the observed in-degrees and out-degrees of $G_0$ 
with their averages $\langle k_i^{in}\rangle_{\Omega}$ and $\langle k_i^{out}\rangle_{\Omega}$ computed on the FiCM ensemble $\Omega$. 
Figure \ref{fig:test_fit} shows such comparison when the average degrees are obtained through eq.~(\ref{eq:prob linking z dir}) for a fully informed FiCM, {\em i.e.}, 
with the value of $z$ computed via eq.~(\ref{eq:estimatez dir}) using the knowledge of in- and out-degrees for all nodes. 
We indeed observe a remarkable agreement between these quantities for our empirical networks: the real degrees are scattered around the functional form of their expected values. 
The amount of deviations from perfect correlation (which would correspond to an actual realization of the FiCM) gives an indication of how well our model describes the real network. 
Note that the validity of hypothesis II can be evaluated also in the case of partial information by performing such comparison on the subset $I$ of nodes whose topological properties are available.

\begin{figure}[p]
\centering
\includegraphics[width=0.85\textwidth]{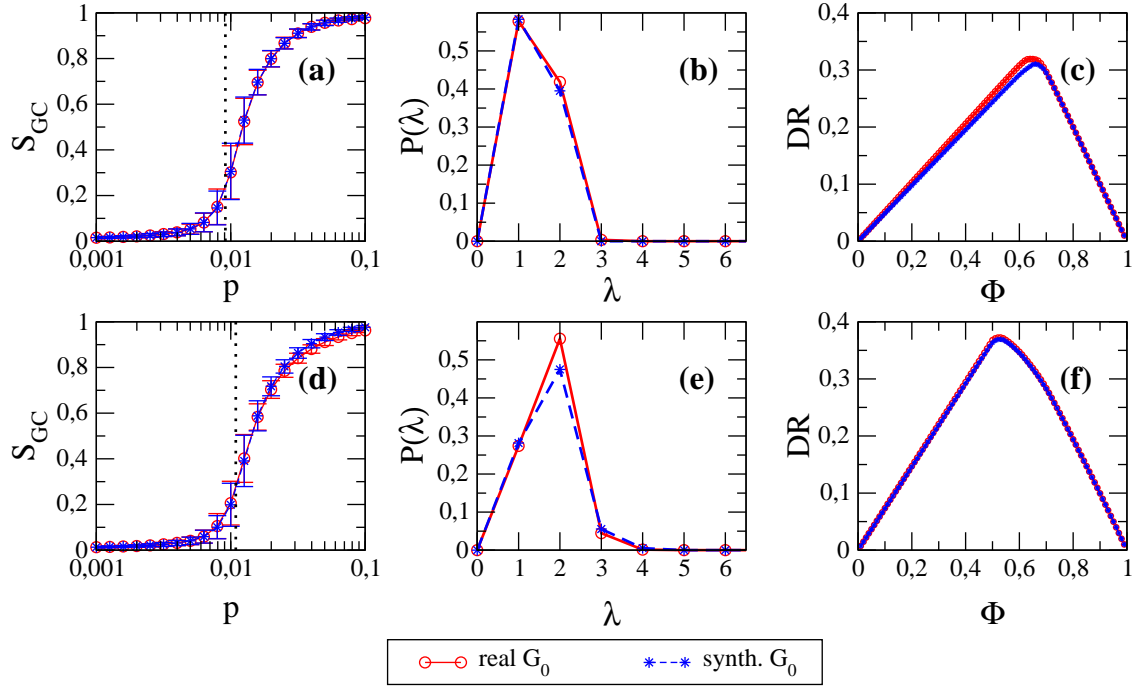}
\caption{Properties of real and synthetic networks. 
Left plots (a,d): dependence of the size of the giant component $S_{GC}$ on the occupation probability $p$ (the vertical dotted line indicates $p^*$). 
Central plots (b,e): probability distribution of the directed shortest path length $\lambda$. 
Right plots (c,f): dependence of $DR$ on the initial distress $\Phi$. 
Top panels (a,b,c) refer to WTW, bottom panels (d,e,f) to E-mid. 
Correlation coefficient values $c$ between real and synthetic curves: (a) $c=0.999$, (b) $c=0.999$, (c) $c=0.994$, (d) $c=0.999$, (e) $c=0.989$, (f) $c=0.998$.}
\label{fig:test_stat}
\end{figure}

In the following, in order to have a quantitative global assessments of the errors caused by hypothesis II, we will test our network reconstruction method both on real networks 
and on benchmark synthetic networks numerically generated with the fully informed FiCM through eq.~(\ref{eq:prob linking z dir}). 
In the latter case, the errors made by the method will be due only to the limited information available about the degree sequences. 
It is then interesting to check whether such generated synthetic networks are equivalent to the real networks in term of systemic risk. 
Figure \ref{fig:test_stat} shows that bond percolation properties, shortest path length distribution and DebtRank values of synthetic networks are in excellent agreement with those of their real counterparts 
(the correlation coefficients between real and synthetic curves are all above 0.99). FiCM thus proves itself to be a proper framework for modeling our empirical networks.

\subsection*{Test against limited information}

In this section we finally proceed to the key testing of the method against the third (and more relevant) source of errors: 
the limitedness of the information available on the degree sequences for calibrating the FiCM. 
In order to obtain a quantitative estimation of the method effectiveness in reconstructing a topological property $X$ of a given a network $G_0$ (which can be either the real one or its synthetic version), 
we implement a procedure consisting in the following operative steps:
\begin{itemize}
\item Choose a value of $n<N$ (the number of nodes for which the in- and out- degrees are known).
\item Build a set of $M=100$ subsets $\{I_\alpha\}_{\alpha=1}^M$ of $n$ nodes picked at random from $G_0$.
\item For each subset $I_\alpha$, use the degree sequences from $G_0$ to evaluate $z$ from eq.~(\ref{eq:estimatez dir}), and name such value $z_\alpha$.
\item Build the ensemble $\Omega(z_\alpha)$ using the linking probabilities from eq.~(\ref{eq:prob linking z dir}): 
generate $m=100$ networks from $\Omega(z_\alpha)$, and compute the average value $X_\alpha$ of property $X$ on this ensemble.
\item Compute the relative root mean square error (rRMSE) of property $X$ over the subsets $\{I_\alpha\}_{\alpha=1}^M$:
\begin{equation}\label{eq.RMSE}
\rho[X]=\mbox{rRMSE}_X\equiv\sqrt{\frac{1}{M}\sum_{\alpha=1}^M\left[\frac{X_\alpha}{X(G_0)}-1\right]^2}
\end{equation}
where $X(G_0)$ is the value of $X$ measured on $G_0$.
\end{itemize}

\begin{figure}[p]
\centering
\includegraphics[width=0.9\textwidth]{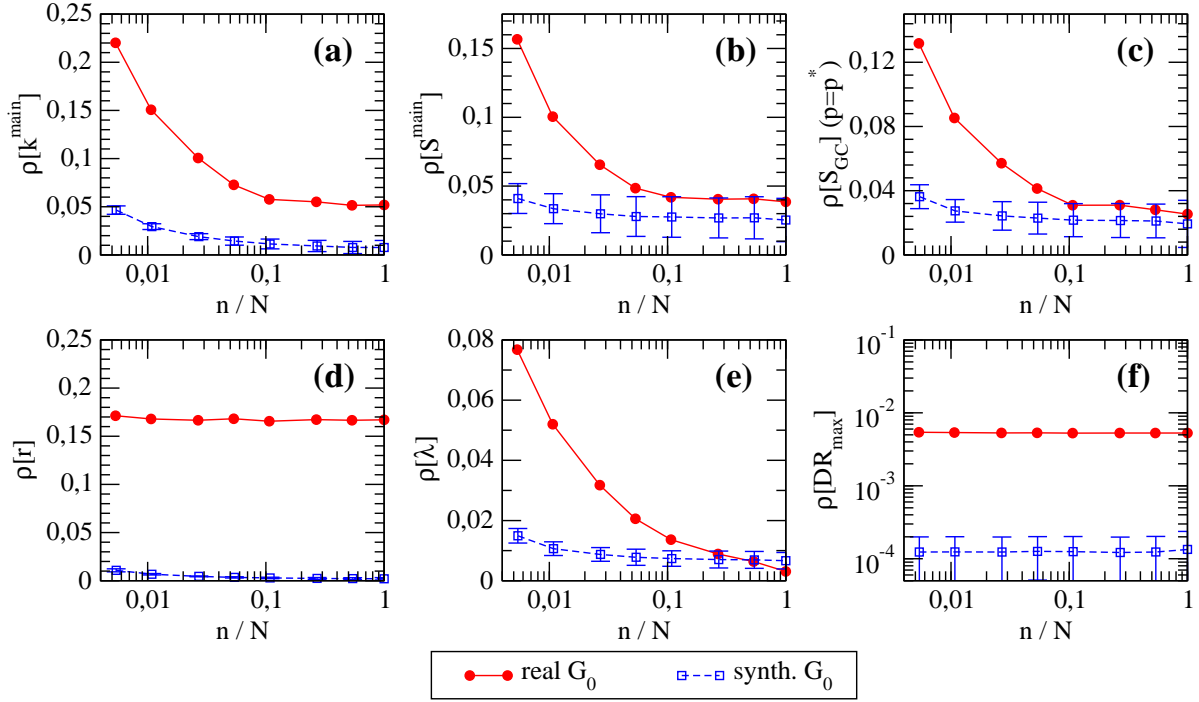}
\caption{rRMSE of various topological properties versus $n$ for the reconstructed $G_0$ of the WTW (both the real network and its synthetic version). 
rRMSE for: (a) degree of the main core $k^{main}$, (b) size of the main core $S^{main}$, (c) size of the giant component $S_{GC}$ at the bond percolation threshold $p^*=\bar{k}^{-1}$, 
(d) link reciprocity $r$, (e) mean shortest path length $\lambda$, (f) maximum value of the group DebtRank $DR$.}
\label{fig:tests_wtw}
\end{figure}

\begin{figure}[p]
\centering
\includegraphics[width=0.9\textwidth]{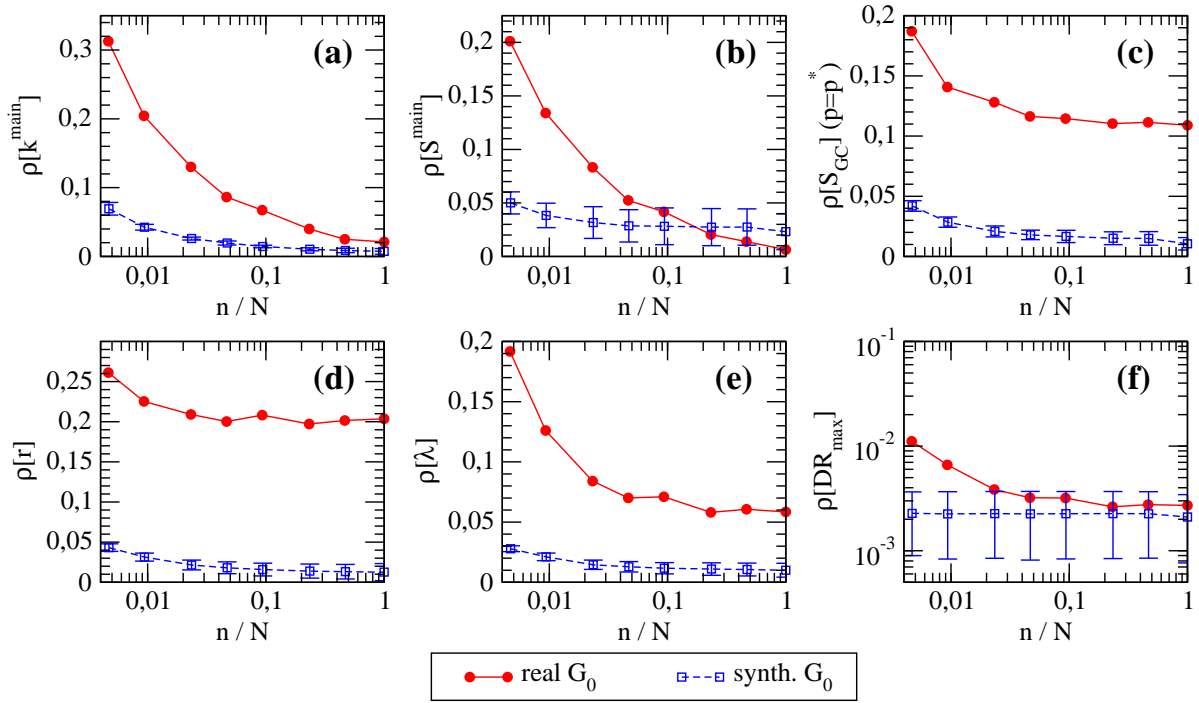}
\caption{rRMSE of various topological properties versus $n$ for the reconstructed $G_0$ of the E-mid (both the real network and its synthetic version). 
rRMSE for: (a) degree of the main core $k^{main}$, (b) size of the main core $S^{main}$, (c) size of the giant component $S_{GC}$ at the bond percolation threshold $p^*=\bar{k}^{-1}$, 
(d) link reciprocity $r$, (e) mean shortest path length $\lambda$, (f) maximum value of the group DebtRank $DR$.}
\label{fig:tests_emid}
\end{figure}

We then study how the rRMSE for the various network properties we consider varies as a function of the size $n$ of the subset of nodes used to calibrate the FiCM 
(\emph{i.e.}, for which in- and out- degree information is available). Results are shown in Figures \ref{fig:tests_wtw} and \ref{fig:tests_emid}. 
We observe that in most of the cases there is a rapid decrease of the relative error as the number of nodes $n$ used to reconstruct the topology increases. 
For instance, generally the error drops to half of the starting rRMSE (for $n=1$) at $n/N=5\%$, and to one quarter for $n/N=10\%$---a value that is rather close to that of the final error made at $n\equiv N$.
This is an indication of the goodness of the estimation provided by our method. As expected, the rRMSE is higher for real networks than for synthetic networks, 
and the difference between the two curves gives a quantitative estimation of the error made in modeling real networks with the FiCM. 
The fact that such a difference is higher for E-mid than for WTW is directly related to a slightly better correlation between real and expected degrees observed in the latter case (Figure \ref{fig:test_fit}). 
Note that the various rRMSE for synthetic networks do not necessarily tend to zero, because the generated synthetic configuration might be highly improbable---in some cases, 
the synthetic network can be even more atypical than the real one. We thus indicate with error bars the range of performance of our method for different choices of synthetic $G_0$. 

Generally, $S_{GC}$, $\lambda$, $k_{main}$ and $S_{main}$ are the properties which are reconstructed better: for instance, with the knowledge of only 10\% of the nodes, 
all the relative errors become smaller than 10\%, and they decrease for increasing $n$. The rRMSE for $r$ and $DR$ show instead a behavior almost flat in $n$. 
The fact that the rRMSE for $r$ computed for real networks remains steadily high is probably due to the fact that reciprocity is hardly reproduced by a directed CM, 
and is better suited as additional imposed constraint \cite{Squartini2012}. The rRMSE for $DR$ is instead remarkably small for real networks (with values around 0.5\%), 
and we can thus conclude that our method is efficient in estimating $DR$ also when the available information is minimal. 
This is particularly relevant to our analysis, since we are estimating $DR$ at its peak ({\em i.e.}, at its maximum, and thus mostly fluctuating, value), 
where the details of the weighted topology play a fundamental role in the process of risk propagation. 
Besides, and more importantly, the value of $DR$ for the real network is computed using the original weighted topology, whereas, 
the computation of $DR$ in the reconstructed network builds on link weights obtained by the simple prescription of eq.~(\ref{eq:weight z dir}). 

In conclusion, the outcome of this analysis is that our network reconstruction method is able to estimate the network properties related to systemic risk with good approximation, 
by using the information on the number of connections of a relatively small fraction of nodes---as long as the fitnesses of all nodes is known.

\section*{Discussion}

In this paper we studied a novel method that allows to reconstruct a directed weighted network and estimate its topological properties by using only partial information 
about its connection patterns, as well as two additional intrinsic properties (interpreted as fitnesses) associated to each node. Tests on empirical networks as well as on synthetic networks 
generated through a fitness-induced configuration model reveals that the method is highly valuable in overcoming the lack of topological information that often hinders the estimation of systemic risk 
in economic and financial systems. Indeed, the information exploited by the method is minimal but is (or should be) publicly available for these kind of systems.

Our work originates from the study of Musmeci {\em et al.} \cite{Musmeci2013}, that represented a first attempt in tackling the problem of network reconstruction from partial information 
within the framework of fitness-induced configuration models. Here however we make fundamental improvements to the method, the key advance being that of extending it to directed weighted networks 
(the most general class of networks). In the present form, the method is then suited to reconstruct high-order network properties related to systemic risk, 
a task of primary practical importance the method was conceived to address---that was however far beyond the reach of its original version. 
Besides, the validation of the fitness-induced configuration model approach to model real networks, as well as the reconstruction of benchmark synthetic networks generated as fitness-based counterparts 
of the empirical networks, are both novel ingredients that allow to assess quantitatively the accuracy of the method. Last but not least, the extensive analysis of different temporal snapshots 
of the real networks we provide in the Supplementary Information allows to strengthens considerably the effectiveness and robustness of our method.

We remark that the method we are proposing here, by reproducing both the binary and weighted topology of the network, represent a substantial step forward in the field of network reconstruction. 
In fact, most of the previous works \cite{Wells2004,Lelyveld2006,Degryse2007,Mistrulli2011,Squartini2011} focused on reproducing the strengths of the real network to the detriment of connection patters, 
whereas, only recently it has been realized that a successful reconstruction procedure must resort also on topological constrains \cite{Mastromatteo2012,Musmeci2013}. 
Here we are proposing a method that allows to always reproduce the strengths, but also to tune the network topology through appropriate connection probabilities. 
In this respect, the use of probabilities derived from degree constraints represent the most general case, which include as specific instances both the dense reconstruction techniques 
\cite{Wells2004,Lelyveld2006,Degryse2007,Mistrulli2011} (obtained for $p_{i\rightarrow j}=1$ $\forall i\neq j\in V$) and the sparse reconstruction method \cite{Mastromatteo2012} 
(roughly obtained for $p_{i\rightarrow j}=1-\lambda$ $\forall i\neq j\in V$). 

Note that one should not be much surprised that the knowledge of a small number of nodes allows to precisely estimate a wide range of network properties, 
because the method assumes the additional knowledge of the fitness parameters for all the nodes.
Besides, the effectiveness of the method strongly depends on the accuracy of the fitness model used to calibrate the CM in order to fit the empirical dataset. 
In the case of WTW and E-mid, the fitness model well describes how links are established across nodes, and our method is thus effective in reconstructing the network properties. 
Finally, we remark that the issue of having limited information on the system under investigation, while being typical for social, economic and financial systems (that are privacy-protected), 
is very relevant also for biological systems such as ecological networks, metabolic networks and functional brain networks---where, due to observational limitations and high experimental costs 
for collecting data, detailed topological information about connections is often missing. Notably, our method can be used to reconstruct any network representing a set of 
(directed and weighted) dependencies among the constituents of a complex system, and we thus believe it will find wide applicability in the field of complex networks and statistical physics of networks.

\section*{Acknowledgements}
This work was supported by the EU project GROWTHCOM (611272), the Italian PNR project CRISIS-Lab, the EU project MULTIPLEX (317532) and the Netherlands Organization for Scientific Research (NWO/OCW).
DG acknowledges support from the Dutch Econophysics Foundation (Stichting Econophysics, Leiden, the Netherlands) with funds from beneficiaries of Duyfken Trading Knowledge BV (Amsterdam, the Netherlands). 
We thank Guido Caldarelli and Stefano Battiston for useful discussion.

\end{document}